\documentclass[12pt,amsmath,pra,showpacs,superscriptaddress,subeqn]{revtex4}

\usepackage{amsmath}
\usepackage{amsfonts}
\usepackage{graphicx}
\usepackage{color}
\renewcommand\Im{\operatorname{Im}}

\begin{document}

\title{Numerical studies of the interaction of an atomic sample with the electromagnetic field in two dimensions}
\date{\today}
\author{Maxim Sukharev}
\email{maxim.sukharev@asu.edu}
\affiliation{Department of Applied Sciences and Mathematics, 
Arizona State University, Mesa, Arizona 85212, USA}
\author{Abraham Nitzan}
\email{nitzan@post.tau.ac.il}
\affiliation{School of Chemistry, Tel Aviv University, Tel Aviv, 69978, Israel}

\begin{abstract}
We consider the interaction of electromagnetic radiation of arbitrary polarization with multi-level atoms in a self-consistent manner, taking into account both spatial and temporal dependencies of local fields. This is done by numerically solving the corresponding system of coupled Maxwell-Liouville equations for various geometries. In particular, we scrutinize linear optical properties of nanoscale atomic clusters, demonstrating the significant role played by collective effects and dephasing. It is shown that subwavelength atomic clusters exhibit two resonant modes, one of which is localized slightly below the atomic transition frequency of an individual atom, while the other is positioned considerably above it. As an initial exploration of future applications of this approach, the optical response of core-shell nanostructures, with a core consisting of silver and shell composed of resonant atoms, is examined. 
\end{abstract}


\maketitle

\section{Introduction}\label{intro}
Nanoscale optical materials have long been attracting considerable attention due to many important applications ranging from optical nanodevices \cite{TangNature08}, plasmonic circuitry \cite{BozhevolnyiPhysToday08}, nanoscale sources of coherent radiation \cite{ZhangNature09}, single atom/molecule manipulation \cite{ChuScience91}, bio-medical applications \cite{HomolaABC03,El-SayedNanoLett05}, and many others. Among such exciting applications lies the yet-to-be explored sub-field of nanoscale optical atomic and molecular physics that deals with ensembles of atoms or molecules interacting with dielectric nanoparticles (NP) and their assemblies. Such systems are characterized by a significant spatial dependence of evanescent electromagnetic (EM) fields on the dielectric environment, providing the means to control the behavior of molecular systems by the combination of large EM fields and large field gradients, exemplified by recent work \cite{AbajoOptExpr06,LakowiczNanoLett07,MiddendorfApplSpectrosc08,XiaNanoLett09,VanDuyneJACS10,XiaJPhysChemLett10} on the EM field associated with metal NP dimers and their dependence on particle sizes and interparticle distance. At their core, these phenomena rely mostly on the excitations of surface plasmon-polariton (SPP) resonances \cite{BarnesReview07} in systems comprising metal NPs and their arrays, \cite{SchatzJPCB03} as well as other nanoscale metal surfaces, such as subwavelength diffraction gratings, whose optical properties depend sensitively on their surface topology and material parameters \cite{SukharevJCP09}. Their studies have lead to many applications such as coherent EM energy transport in space \cite{AtwaterAdvMat01}, surface enhanced Raman spectroscopy (SERS) \cite{EtchegoinBook09} and tip-enhanced microscopy \cite{SukharevJPCA09}. Recent attention has focused on optical control scenarios, ranging from coupled exciton-plasmon dynamics in semiconductor nanodots \cite{Woggon07,Sham07,Rogach08,Minami09,Davis10,Wang10,Achermann10,Lodahl10,Gray11} and in molecular aggregates \cite{Reinhoudt02,Plenet06,Rothberg06,Asahi07,Wang08,Ebbesen09,Richards09,Fujii10,Patra10} where metal NPs affect excitation energy transfer between molecules, to optical trapping of single atoms or molecules \cite{ZhanPRA07,RauschenbeutelPRL10,ZhanOpLett10,HakutaPRA10}. Such applications are facilitated by the possibility to control the geometry of nano-materials (NPs sizes, their relative arrangement, etc.) with an outstanding precision \cite{XiaScience02}.

While theoretical and computational methodologies for studying these phenomena have advanced considerably, the consequences of mutual feedback between molecular excitations and metallic SPP resonances are not well understood, especially when one probes systems comprising both metallic nanostructures and semiconductor or molecular particles or layers. An often used simple description is based on assigning a dielectric response function to the semiconductor/molecular component and solve the electromagnetic problem for the corresponding composite dielectric. While such an approach can be useful for describing the effect of the molecular environment on the metallic plasmon, it cannot be used to describe energy-transfer, relaxation, and spontaneous emission in the excitonic (molecular or semiconductor) system. It is hence important to develop a self-consistent description of the electromagnetic response of such systems. Such approach has to take into account the electrodynamics of the radiation field and the quantum dynamics of the molecular system in a self-consistent manner. This can be accomplished by solving simultaneously Maxwell$'$s equations for the radiation field and the Liouville equation for the molecular density matrix, including the molecular polarization current in the former and the molecules-field interaction in the latter.

First attempts to consider numerically coupled Maxwell-Bloch equations have been initiated by Ziolkowski et al. \cite{ZiolkowskiPRA95, ZiolkowskiPRA02} for simple two-level atoms in one and two dimensions utilizing finite-difference time-domain (FDTD) technique. Later on this approach has been extended to three dimensions \cite{RuoccoPRA08}. Although these works contain interesting and  important physics, they are limited to ensembles of two-level systems. Consideration of multilevel systems is critical for modeling of nano-lasing, which has to include at least three levels. Moreover, the proposed numerical implementation results in noticeably, long execution times. The scheme we propose is more efficient as discussed below. Similarly Neuhauser et al. proposed another approach \cite{NeuhauserJCP09} where the authors coupled Maxwell$'$s equations to the Schr$\ddot{o}$dinger equation describing a molecule located in the closed proximity of a metal NP. These works, however, cannot in their present form include relaxation and dephasing effects, which, as we demonstrate, are very important.

In this paper we describe a numerical implementation of such a model, using the methodology developed by Ziolkowski et al. \cite{ZiolkowskiPRA95, ZiolkowskiPRA02} as our starting point. This model captures collective effects that play pivotal role in electrodynamics of nano-systems, as well as the counterbalancing effect of dephasing processes. The questions to be addressed are:
\begin{enumerate}
\item How does a size of the system affect scattering/absorption of EM radiation?
\item What is a role of dephasing and relaxation effects?
\item When does one observe collective response of atoms to external EM excitation?
\end{enumerate}
These and other closely related questions are not only important from the fundamental point of view (how optically induced interatomic or intermolecular interactions depend on structural/material parameters), but also essential for general understanding of optics of many-body systems.

In this regard it should be pointed out that although the technique we propose in this paper is utilized to capture collective effects of quantum particles in the linear response regime, it can easily be applied to nonlinear systems. 

The paper is organized as follows: Section \ref{model} discusses our computational approach, based on coupled Maxwell-Liouville equations in the mean-filed approximation. In Section \ref{fdtd} we provide details of the numerical implementation; Section \ref{results} describes and discusses the results of our numerical studies. Our main conclusions along with future research outlook are presented in Section \ref{conclusion}.

\section{Model}
\label{model}
We consider a general problem of a system of quantum particles (further referred to as atoms) interacting with EM radiation. We start from the time domain Maxwell$'$s equations for the dynamics of the EM fields, $\vec{E}$ and $\vec{H}$
\begin{subequations}
\label{Maxwell}
\begin{equation}
\mu_0\frac{\partial\vec{H}}{\partial t}=-\nabla\times\vec{E},
\label{Maxwell1}
\end{equation}
\begin{equation}
\varepsilon_0\frac{\partial\vec{E}}{\partial t}=\nabla\times\vec{H}-\vec{J},
\label{Maxwell2}
\end{equation}
\end{subequations}
where $\mu_0$ and $\varepsilon_0$ are magnetic permeability and dielectric permittivity of free space, respectively. In spatial regions occupied by a metal nanostructure (such as metal NP, for instance) the equation (\ref{Maxwell2}) is evaluated in the standard way from the metal dielectric dispersion \cite{Gray03}. In the present study the dispersion of dielectric constant of 
metal, $\varepsilon(\omega)$, is taken in the form of the Drude model
\begin{equation}
\label{Drude model}
\varepsilon(\omega)=\varepsilon_r-\frac{\omega^{2}_{p}}{\omega^{2}-i \gamma \omega}
\end{equation}
with numerical parameters describing silver for the wavelengths of interest 
$\varepsilon_r=8.26$, $\omega_p=1.76\times10^{16}$ rad/sec, 
$\gamma=3.08\times10^{14}$ rad/sec. The time evolution of the current density $\vec{J}$ in metal regions is then
\begin{equation}
\label{current}
 \frac{\partial\vec{J}}{\partial t}=a\vec{J}+b\vec{E},
\end{equation}
where $a=-\gamma$ and $b=\varepsilon_0 \omega_p^2$.

In the spatial regions occupied by atoms, the mutual interaction between the atomic system and the EM field is accounted for in a self-consistent manner as follows: first, the current density in Eq. (\ref{Maxwell2}) is expressed in terms of the macroscopic polarization of the atomic system, $\vec{P}\left(\vec{r},t \right)$
\begin{equation}
\label{P_J}
\vec{J}=\frac{\partial\vec{P}}{\partial t}.
\end{equation}
The latter is given by
\begin{equation}
\label{Polarization}
\vec{P}=n_a \langle \vec{\mu}\rangle,
\end{equation}
where
\begin{equation}
\label{mu}
 \langle \vec{\mu}\rangle=Tr(\hat{\rho}\vec{\mu})
 \end{equation}
 is the expectation value of the atomic dipole moment and $n_a$ is the atomic density. Eqs. (\ref{P_J}) - (\ref{mu}) constitutes the main approximation of the present approach, whereupon the local polarizability is expressed in terms of the local atomic density multiplied by the local averaged single atomic dipole. The time evolution of the latter is obtained from the evolution of the single atom density matrix (described below) in the presence of the EM field, thus providing a self-consistent description of the field-matter dynamics.

\begin{figure}[tbph]
\centering\includegraphics[width=\linewidth]{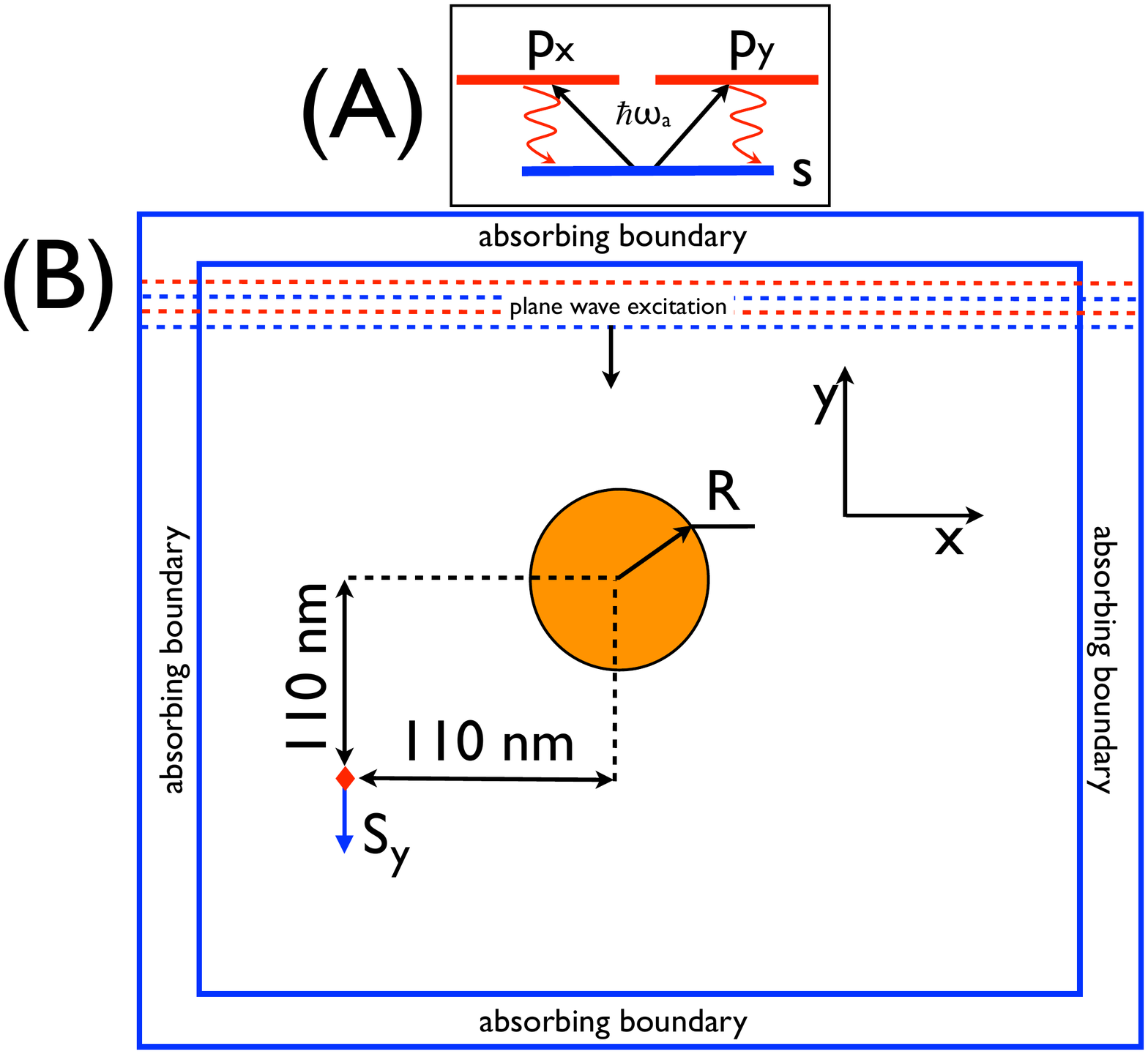}
\caption{(Color online). Panel A shows the energy level diagram of a two-level two-dimensional atom with black arrows indicating optically induced transitions by the TE mode and red arrows representing spontaneous decay. Panel B depicts schematics of the simulations with the detection point shown as a red diamond (in the lower left corner), where the $y$-component of the Poynting vector is calculated.}
\label{setup}
\end{figure}

Next consider the atomic subsystem. While our ultimate goal is to study realistic 3-dimensional systems, the present study focuses on nanoscale atomic clusters in two dimensions, taken to lie in the XY plane \cite{density note}. The incident radiation field is represented by a transverse-electric (TE) mode with respect to $z$-axis. It is characterized by two in-plane electric field components, $E_x$ and $E_y$, and one out-of-plane magnetic field component, $H_z$. To account for the (two-dimensional) spherical symmetry of the atomic polarization response, the atoms are described as 3-level systems: an $s$-type ground state and two degenerate $p$-type excited states of $p_x$ and $p_y$ character (as depicted in Fig. \ref{setup}A). In anticipation of possible generalizations to more complex models involving multilevel systems we use, in what follows, a basis of angular momentum wavefunctions with quantization axis in the $z$ direction, $\left|1\right>=\left|s\right>$, $\left|2\right>=\left( \left|p_x\right>+i \left|p_y\right> \right)/\sqrt2$, $\left|3\right>=\left( \left|p_x\right>-i \left|p_y\right> \right)/\sqrt2$, with optical transitions corresponding to $\Delta J=\pm 1$ and $\Delta M=\pm 1$ selection rules. The corresponding Hamiltonian is
\begin{equation}
\label{Hamiltonian}
\hat{H}=\hat{H_0}-\vec{\mu}\cdot\vec{E}(t)=
\left( 
\begin{array}{ccc}
0 & \Omega_-\left(t \right) & -\Omega_+\left(t \right) \\
\Omega_+\left(t \right) & \hbar\omega_a & 0 \\
-\Omega_-\left(t \right) & 0 & \hbar\omega_a
\end{array} 
\right),
\end{equation}
where $\hbar\omega_a$ is the atomic energy transition, $\Omega_\pm=\mu_{sp}\left(E_x\left(t \right)\pm iE_y\left(t \right)\right) / \sqrt6$, and $\mu_{sp}$ is $s-p$ the matrix element of the dipole moment operator, that, in principle, can be taken from experiment or calculated using standard quantum chemistry packages. Its Cartesian components are
\begin{subequations}
\label{dipole moment}
\begin{align}
\hat{\mu}_x=-\frac{\partial \hat{H}}{\partial E_x}=\frac{\mu_{sp}}{\sqrt6}
\left( 
\begin{array}{ccc}
0 & -1 & 1 \\
-1 & 0 & 0 \\
1 & 0 & 0
\end{array} 
\right),
\\
\hat{\mu}_y=-\frac{\partial \hat{H}}{\partial E_y}=\frac{\mu_{sp}}{\sqrt6}
\left( 
\begin{array}{ccc}
0 & i & i \\
-i & 0 & 0 \\
-i & 0 & 0
\end{array} 
\right).
\end{align}
\end{subequations}
Note that the dipole moment operator (\ref{dipole moment}) differs from the one used in \cite{ZiolkowskiPRA02} by the factor of $\sqrt 3$.

The mean field approximation Eq. (\ref{Polarization}) make it possible to describe the atomic system in terms of the single atom density matrix $\hat{\rho}$, which satisfies the Liouville equation
\begin{equation}
\label{Liouville}
i\hbar\frac{d\hat{\rho}}{dt}=[\hat{H},\hat{\rho}]-i\hbar\hat{\Gamma}\hat{\rho}.
\end{equation}
Eq. (\ref{Liouville}) describes the time evolution of an atom interacting with the radiation field and subject to (assumed Markovian) relaxation processes described by the $\hat{\Gamma}$ operator, which is taken in the Lindblad form \cite{PetruccioneBook02}. The diagonal elements of this operator correspond to excited states lifetimes, while nondiagonal elements account for dephasing effects.

Eqs. (\ref{Maxwell}) - (\ref{Liouville}) describe the time evolution of the atomic system and radiation field in a self-consistent way. Note that in the single-atom Hamiltonian $\hat{H}$ given by Eq. (\ref{Hamiltonian}) interatomic interactions are absent. Still, the dynamics described by Eqs. (\ref{Maxwell}) - (\ref{Liouville}) is not that of independent atoms: the self-consistent scheme accounts for all interactions between atoms that are associated with their mutual interaction with the radiation field, including excitonic (energy transfer) interactions, which are all important for elucidating the overall system response. Note that, assuming that the EM field does not significantly vary within a volume occupied by a single atom, the spatial dependence of the density matrix in Eq. (\ref{Liouville}) depends  parametrically on position via the EM field variables.

Eqs. (\ref{Liouville})-(\ref{dipole moment}) lead to the following equations for the atomic density matrix elements
\begin{subequations}
\label{density matrix equations}
\begin{equation}
\frac{d\rho_{11}}{dt}=i\omega_{+}\left( \rho_{12}+\rho_{13}^{*}\right)-i\omega_{-}\left( \rho_{13}+\rho_{12}^{*}\right)+\gamma_1\left( \rho_{22}+\rho_{33}\right),
\label{density matrix equations1}
\end{equation}
\begin{equation}
\frac{d\rho_{12}}{dt}=i\omega_a\rho_{12}-i\omega_{-}\left( \rho_{22}-\rho_{11}\right)+i\omega_{+}\rho_{23}^{*}-\gamma_2\rho_{12},
\label{density matrix equations2}
\end{equation}
\begin{equation}
\frac{d\rho_{13}}{dt}=i\omega_a\rho_{13}+i\omega_{+}\left( \rho_{33}-\rho_{11}\right)-i\omega_{-}\rho_{23}-\gamma_2\rho_{13},
\label{density matrix equations3}
\end{equation}
\begin{equation}
\frac{d\rho_{22}}{dt}=i\omega_{-}\rho_{12}^{*}-i\omega_{+}\rho_{12}-\gamma_1\rho_{22},
\label{density matrix equations4}
\end{equation}
\begin{equation}
\frac{d\rho_{23}}{dt}=-i\omega_{+}\left( \rho_{13}+\rho_{12}^{*}\right)-2\gamma_2\rho_{23},
\label{density matrix equations5}
\end{equation}
\begin{equation}
\frac{d\rho_{33}}{dt}=i\omega_{-}\rho_{13}-i\omega_{+}\rho_{13}^{*}-\gamma_1\rho_{33},
\label{density matrix equations6}
\end{equation}
\end{subequations}
where $\omega_\pm=\Omega_\pm / \hbar$, $\gamma_2=\gamma_p+\gamma_1/2$ with $\gamma_p$ denoting the pure dephasing rate due to environmentally induced random fluctuations in the atomic energy spacing. As noted above, in Eq. (\ref{density matrix equations}) we denoted the ground state as $\left| 1 \right>$ and the excited states $\left| J=1, M=-1 \right>$ and $\left| J=1, M=1 \right>$ as $\left| 2 \right>$ and $\left| 3 \right>$, respectively.

Finally, using Eqs. (\ref{dipole moment}) and (\ref{density matrix equations}) we obtain the macroscopic polarization current (time derivative of Eq. (\ref{Polarization})), which enters Ampere's law (\ref{Maxwell2})
\begin{subequations}
\label{polarization current}
\begin{equation}
\frac{\partial \vec{P}}{\partial t}=n_a \frac{\partial \langle \vec{\mu}\rangle}{\partial t},
\label{polarization current1}
\end{equation}
where, back in cartesian coordinates,
\begin{equation}
\frac{\partial \langle \mu_x \rangle}{\partial t}=\frac{E_y \mu_{sp}^2}{3\hbar}\left( \rho_{22}-\rho_{33}\right)-i\frac{\mu_{sp}}{\sqrt6}\left[ \left( \omega_a+i\gamma_2\right) 
\left( \rho_{12}-\rho_{13}\right)-\left( \omega_a-i\gamma_2\right) \left( \rho_{12}^{*}-\rho_{13}^{*}\right)\right],
\label{polarization current2}
\end{equation}
\begin{equation}
\frac{\partial \langle \mu_y \rangle}{\partial t}=-\frac{E_x \mu_{sp}^2}{3\hbar}\left( \rho_{22}-\rho_{33}\right)+\frac{\mu_{sp}}{\sqrt6}\left[\left( \omega_a+i\gamma_2\right) 
\left( \rho_{12}+\rho_{13}\right)+\left( \omega_a-i\gamma_2\right) \left( \rho_{12}^{*}+\rho_{13}^{*}\right)\right],
\label{polarization current3}
\end{equation}
\end{subequations}

We end this section with two comments. First, as already pointed out, Eqs. (\ref{density matrix equations}) a mean field description of a system of atoms interacting with the EM field. In this approximation a single atom interacts with other atoms through the electromagnetic field associated with their mean local density. Obviously, such an approach cannot account for specific atom-atom correlations, but it can describe collective effects in a system of atoms resulting from their interaction with the EM field.

Second, we note that this procedure can be easily generalized to yield the analogous 3-dimensional coupled Maxwell-Liouville equations. To maintain spherical symmetry we would need to include an additional atomic level $\left| J=1, M=0 \right>$, which is coupled to the ground atomic state by $E_z$ \cite{RuoccoPRA08}. Obviously, it is also possible to expand the atomic basis and consider additional excited manifold starting with $\left| J=2, M \right>$. Although the number of equations similar to Eqs. (\ref{density matrix equations}) grows significantly, modern analytical computer packages such as Mathematica \cite{AMO} can easily handle necessary algebra and subsequent computer coding. For a molecule without rotational symmetry, average over angular distribution in the calculation of $\vec{P}$ from Eq. (\ref{Polarization}) may be used when relevant.

\section{Numerical approach}
\label{fdtd}
To solve the system (\ref{Maxwell}), (\ref{density matrix equations}), (\ref{polarization current}) of coupled Maxwell-Liouville equations we employ a generalized FDTD technique \cite{Taflove}. Within the FDTD, both the electric, $\vec{E}$, and magnetic, $\vec{H}$, fields are propagated in time and space by directly discretizing Maxwell$'$s equations (\ref{Maxwell}). This approach has several attractive technical features, including its numerical stability and the explicit description of the magnetic field. The latter is especially important if one considers structures with sharp corners, at which the tangential components of $\vec{H}$ have singularities. For the our purposes, the main advantage of the FDTD approach is that the boundary conditions (i.e. the continuity of the tangential $\vec{H}$ and $\vec{E}$ components) are automatically maintained at all grid points owing to the use of the Yee cell \cite{Yee}. This allows straightforward programming of the complex geometries.

For simulations of open systems, one needs to impose artificial absorbing boundaries in order to avoid numerical reflection of outgoing EM waves back to the simulation domain. Among the various approaches that address this numerical issue, the perfectly matched layers (PML) technique is considered to be the most powerful \cite{Berenger}. It reduces the reflection coefficient of outgoing waves at the simulation region boundary to $\sim10^{-8}$. In essence, the PML approach surrounds the simulation domain by thin layers of non-physical material that efficiently absorbs outgoing waves incident at any angle. We have implemented the most efficient and least memory variant of the method, the convolution perfectly matched layers (CPML) absorbing boundaries \cite{CPML}. Through extensive numerical experimentation, we have empirically determined optimal parameters for the CPML boundaries that lead to almost no reflection of the outgoing EM waves at all incident angles.

In the calculations reported below we consider structures with a characteristic size much smaller than the incident wavelength. Hence it is a good approximation to excite such systems using a plane wave. The latter is accomplished via implementation of total field/scattered field approach \cite{Taflove} within the FDTD.

We partition the FDTD scheme onto an array of parallel grid slices by dividing the cubic simulation cell into M $xy$ slices, where M is the number of available processors. Point-to-point message passing interface (MPI) communication subroutines \cite{MPI} are implemented at the boundaries between slices. The number of $xy$ planes in each slice usually varies in the range from 15 to 20. All simulations are performed on the home-build 128-core AMD Opteron based cluster at Arizona State University \cite{plasmon}.

The numerical implementation of the proposed scheme is as follows
\begin{enumerate}
\item In the spatial regions occupied by atoms the Maxwell equations are solved utilizing the standard FDTD algorithm. First, magnetic field is updated according to Faraday$'$s law, Eq. (\ref{Maxwell1}). Next, using Ampere$'$s law, Eq. (\ref{Maxwell2}), we update the electric field with the macroscopic polarization current density, Eqs. (\ref{polarization current}), which is calculated using the density matrix of the previous time step. The EM fields in the regions occupied by metal are updated according to the auxiliary differential equation method, \cite{Taflove} Eq. (\ref{current}).

\item With the knowledge of local electric field components (stored in memory at two previous time steps) we update the density matrix at each spatial point on the grid according to Eq. (\ref{density matrix equations}) using the fourth order Runge-Kutta scheme \cite{Runge-Kutta}.

\item Finally, with the knowledge of the electric field components and the updated density matrix we calculate the macroscopic polarization current, $\frac{\partial\vec{P}}{\partial t}$, at each grid point according to (\ref{polarization current}). 
\end{enumerate}
We have verified this scheme using several test cases. A most important test of numerical stability is to check that the condition Tr$(\hat{\rho})=1$ is maintained at each time step. In all simulations this condition was perfectly satisfied with almost no dependence on incident field amplitude and other physical parameters. Another interesting test was to demonstrate the absence of self-interactions in our calculation. Such interactions often appear spuriously in mean field calculations, whereupon a particle interacts with its own contribution to the mean density. In the present situation, however, the field produced by the oscillating dipole of a given atom propagates away from this atom and can affect it only through the polarization induced in other atoms or (in different setups) through reflection from the boundaries, both physically valid phenomena. We have verified that direct self-interaction is indeed absent in our calculation by solving Eqs. (\ref{Maxwell}) and (\ref{density matrix equations}) for the case where the system occupies a single grid point: the same solution is obtained whether or not the polarization source term is included in Eq. (\ref{Maxwell2}). Finally, we have compared our results and execution time of the proposed integration scheme with those obtained by Ziolkowski et al. \cite{ZiolkowskiPRA02}. We have implemented the numerical approach based on the predict-corrector method and atomic basis as used in \cite{ZiolkowskiPRA02} and compared it with ours (keeping in mind that $\mu_{sp}$ has to be re-normalized by the factor of $\sqrt3$). The simulation data obtained using both approaches were in excellent agreement. However execution times for the codes employing our approach were noticeably smaller.

With the solution of Eqs. (\ref{Maxwell}), (\ref{density matrix equations}), (\ref{polarization current}) obtained in this way, the following observables can be calculated:
\begin{description}
\item[(1)] The scattered radiation can be computed as the difference between the total and the incident EM fields. At any detection point, e.g. that depicted as a red diamond in Fig. \ref{setup}B, we can calculate the Poynting vector components associated with the scattered EM field. This may be integrated over a spherical boundary surrounding the atomic system to yield the total scattered radiation. These calculations can be accomplished in a transient mode to give the time dependent response to an incident EM pulse, or in a steady state mode that yields the long time steady state response to CW incident radiation of a given frequency, $\omega_{\mbox{in}}$. In the latter case we need to propagate the Maxwell-Liouville equations under the incident CW radiation until steady-state is reached.
\item[(2)] Generally, for a given incident frequency $\omega_{\mbox{in}}$ the outgoing field may exhibit different frequencies $\omega_{\mbox{out}}$ (with amplitudes obtained by Fourier-transforming the scattered signal), making it possible to obtain the outgoing steady state flux (Poynting vector) in a given direction at any $\omega_{\mbox{out}}$ for a given incident frequency. Integrating the outgoing flux over $\omega_{\mbox{out}}$ and displaying the result as a function of $\omega_{\mbox{in}}$ yields the absorption spectrum of the atomic cluster.
\item[(3)] A much easier way to calculate the steady state absorption lineshape is by calculating the steady state relaxation flux of the excited state populations according to 
\begin{equation}
\label{absorption}
-\frac{dE}{dt}\approx\hbar\omega_a\gamma_1\int d^3r\left( \rho_{22}\left( \vec{r}\right)+\rho_{33}\left( \vec{r}\right)\right),
\end{equation}
where the integral is taken over cluster volume \cite{CW volume}. Eq. (\ref{absorption}) expresses for any given incoming frequency the rate of energy dissipation by the molecular system, which at steady state should be equal to the rate of energy absorption at that frequency.
\item[(4)] For some applications the short pulse method (SPM) \cite{Gray03} can save a substantial computing effort. In this approach we use an ultra-short incident pulse with a wide bandwidth, which is almost flat in the spectral region of interest (in our simulations the incident pulse duration was set at 0.36 fs, which corresponds to flat spectrum throughout the frequency domain considered in this paper). Such pulse can be represented as a coherent linear superposition of CW plane waves with different $\omega_{\mbox{in}}$. We then propagate Maxwell-Liouville equations for several picoseconds (the total propagation time has to be significantly longer than the lifetime of excited states of an atom, $1/\gamma_1$), and take the Fourier transform of the calculated field. Under conditions of linear response {\it and} elastic scattering (namely, when $\omega_{\mbox{out}}=\omega_{\mbox{in}}$), the Fourier component at frequency $\omega$ contains all the information relevant to a CW process at frequency $\omega$.
\end{description}

When applicable, the SPM can save substantially on computation time, since it yields the system response at many frequencies from one short time computation. It is important to understand its shortcomings. Two limitations, mentioned above, are obvious: this method is applicable only in linear response and only when the light scattering process is elastic. Both limitations are associated with the requirement that a given incident frequency can give rise to response only at the same frequency. A third limitation is important in the present context, because Eqs. (\ref{density matrix equations2}) and (\ref{density matrix equations3}) are inherently nonlinear. Linearity is obtained when $\rho_{22}$ and $\rho_{33}$ may be neglected, and $\rho_{11}$ taken constant in these equations. In this case, the steady state solutions for $\rho_{12}$ and $\rho_{13}$, and therefore the polarization (Eq. (\ref{Polarization})), oscillate with the incident frequency as required. Care has to be exercised if one attempts to calculate the steady state excitation, $\rho_{22}+\rho_{33}$, in this method. An approximation may be obtained if the incident frequency is close to atomic transition frequency, $\omega_a$, by taking the Fourier transform at $\omega=0$, i.e. the time average, of the resulting time dependent signal.

The method, as described, cannot describe spontaneous emission, i.e. fluorescence, since the latter is a quantum effect associated with the quantum nature of the radiation field. It has been demonstrated \cite{Cao09} that the effect of spontaneous emission can be accounted for partially by imposing a classical stochastic field on the system. Obviously, a classical EM noise cannot mimic vacuum fluctuations; in particular it can induce excitation of ground state molecules while vacuum fluctuations can lead only to radiative damping of excited molecules. One can use this trick to study time evolution in a system which is initially inverted, that is, all molecules in the excited state. In this case, the induction of emission by the EM noise will soon lead to a dominant signal of induced emission that does not depend much on the nature of that noise, provided the latter is weak enough. This method has been succesfully used to simulate superradiance\cite{Cao09,Ho08} and gain \cite{Cao10} within the FDTD approach. However, this method cannot be used to generate fluorescence in a system of mostly ground state molecules ($\rho_{22}+\rho_{33}\ll\rho_{11}$), where the main effect of such noise will be to induce unphysical molecular excitation.

\section{Results and discussion}
\label{results}
The simulation setup is shown in Fig. \ref{setup}B: an atomic cluster is excited with an $x$-polarized low intense plane wave propagating in the $y$ direction. We use low intensity incoming fields (our incident electric field was fixed at 1 V/m) in order to insure linearity of the system response \cite{linear}. Eqs. (\ref{Maxwell}), (\ref{density matrix equations}), (\ref{polarization current}) are then evolved to yield the electric and magnetic fields as functions of time and position. In the current studies we focus on the $y$ -component of the Poynting vector, $S_y\sim E_x H_z$, as the observable of interest.

\begin{figure}[tbph]
\centering\includegraphics[width=\linewidth]{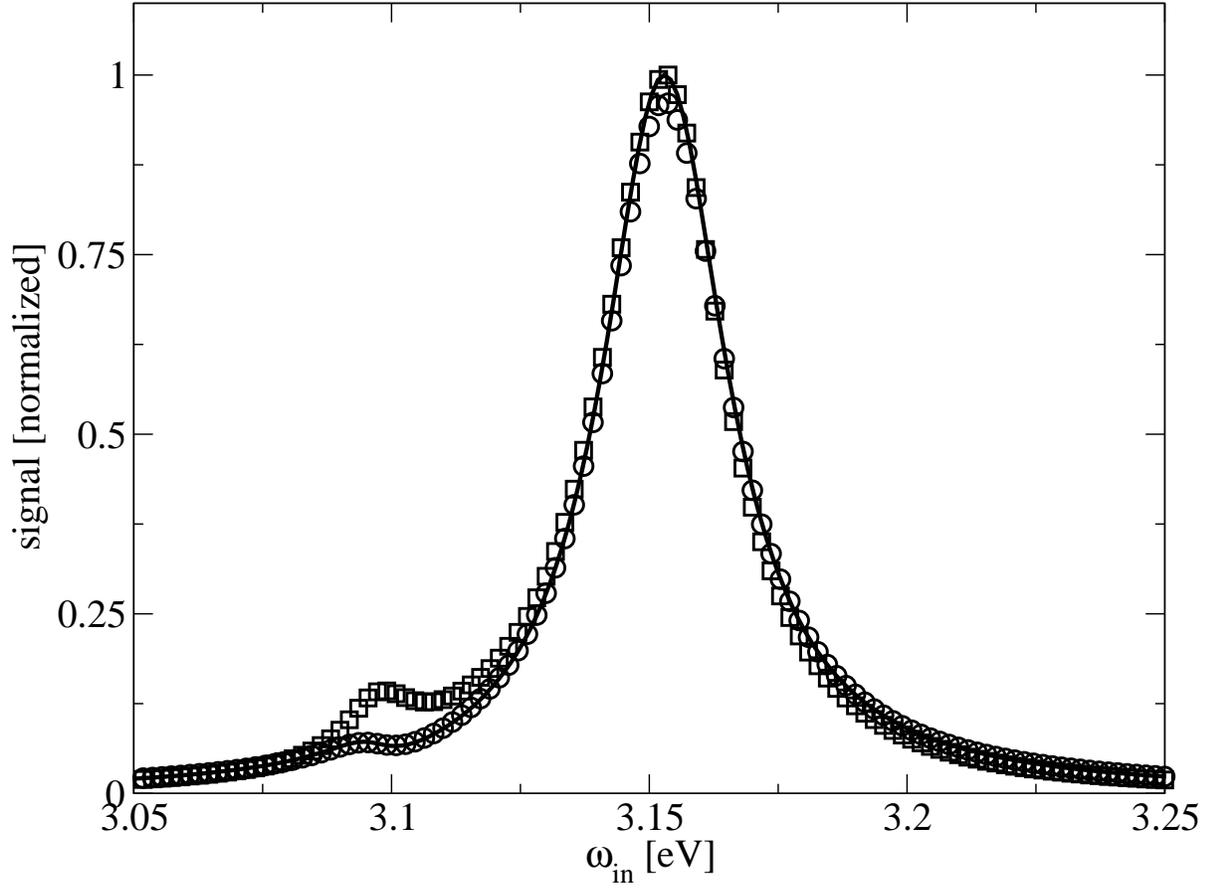}
\caption{Scattering intensity as a function of the incoming frequency $\omega_{\mbox{in}}$ calculated within SPM approach (solid line) and CW scheme (circles). Normalized absorption (see Eq.(\ref{absorption})) is shown as squares. Simulations are performed for the cluster with the following parameters: $\omega_a=3.1$ eV, $R=25$ nm, $n_a=7\times10^{25}$ m$^{-3}$, $\gamma_1=10^{12}$ s$^{-1}$, $\gamma_p=10^{13}$ s$^{-1}$, $\mu_{sp}=25$ Debye.}
\label{CW-SPM}
\end{figure}

Fig. \ref{CW-SPM} presents direct comparison of two methods: SPM approach and CW scheme. The scattering signals obtained from the two methods are in excellent agreement. Also shown is the absorption spectrum from the CW calculation using Eq. (\ref{absorption}). The absorption lineshape (normalized so that it matches the scattering signals at the peak) also leads to nearly identical lineshape, except that it exhibits a more pronounced resonance near the atomic transition frequency. 

It is not surprising, but still providing a consistency check, that at the very low densities ($n_a<10^{24}$ m$^{-3}$) and zero dephasing our simulations are in the perfect agreement with the Clausius-Mossotti approximation described above. Moreover, calculations, in which the polarization current term in (\ref{Maxwell2}) was neglected (i.e. the atoms are not coupled to each other through their mutual interaction with the radiation field but rather driven only by external incident radiation), were in a perfect agreement with the data produced by the full self-consistent computations in the limit of low densities.

\begin{figure}[tbph]
\centering\includegraphics[width=\linewidth]{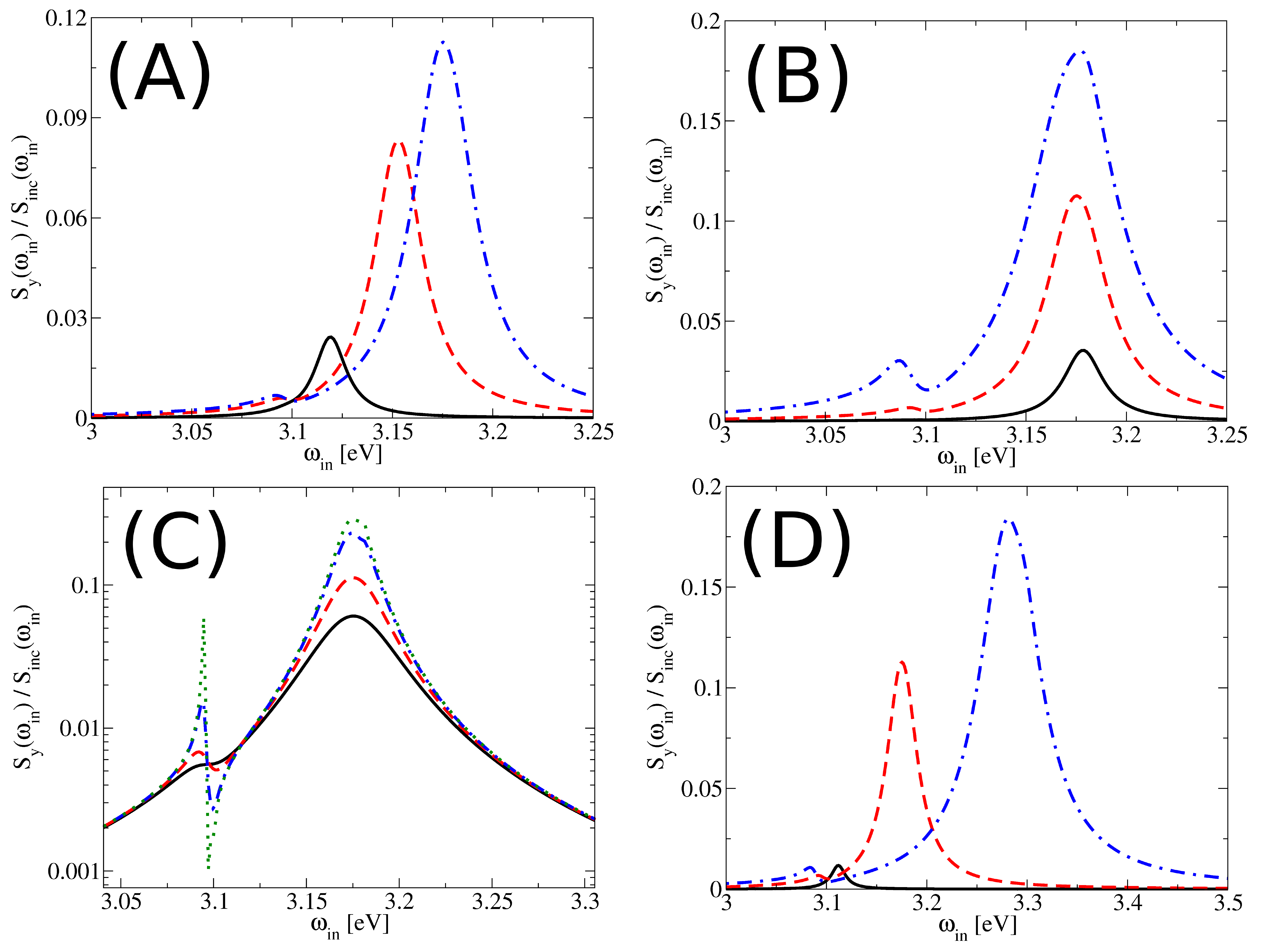}
\caption{(Color online). Linear optics of atomic clusters with $\omega_a=3.1$ eV, $\gamma_1=10^{12}$ s$^{-1}$. 
\textbf{Panel A}: scattering intensity as a function of the incident frequency for three atomic densities, $n_a$ - black solid line: $n_a=2.5\times10^{25}$ m$^{-3}$ (ideal gas at atmospheric pressure and room temperature), red dashed line: $n_a=7\times10^{25}$ m$^{-3}$, and blue dash-dotted line: $n_a=10^{26}$ m$^{-3}$. Other parameters are: $R=25$ nm, $\gamma_p=10^{13}$ s$^{-1}$, $\mu_{sp}=25$ Debye. 
\textbf{Panel B}: same is in panel A for three cluster's radii, $R$ - black (solid), red (dashed), and blue (dash-dotted) lines show scattering intensity results for $R=15$ nm, $25$ nm, and $35$ nm, respectively. Other parameters are: $n_a=10^{26}$ m$^{-3}$, $\gamma_p=10^{13}$ s$^{-1}$, $\mu_{sp}=25$ Debye. 
\textbf{Panel C}: same as in panels A and B (now shown in logarithmic scale), but for four pure dephasing rates, $\gamma_p$ - black (solid) line: $\gamma_p=2\times10^{13}$ s$^{-1}$, red (dashed) line: $1\gamma_p=10^{13}$ s$^{-1}$, blue (dash-dotted) line $\gamma_p=2\times10^{12}$ s$^{-1}$, and green (dotted) line shows the data without pure dephasing $\gamma_p=0$ s$^{-1}$. Other parameters are: $R=25$ nm, $n_a=10^{26}$ m$^{-3}$, $\mu_{sp}=25$ Debye. 
\textbf{Panel D}: same as in panels A-C, but for three values of the matrix element of the dipole moment, $\mu_{sp}$ - black (solid), red (dashed), and blue (dash-dotted) lines correspond to atomic systems characterized by $\mu_{sp}=10$ Debye, $\mu_{sp}=25$ Debye, and $\mu_{sp}=40$ Debye, respectively. Other parameters are: $R=25$ nm, $n_a=10^{26}$ m$^{-3}$, $\gamma_1=10^{12}$ s$^{-1}$, $\gamma_p=10^{13}$ s$^{-1}$.}
\label{atoms}
\end{figure}

\begin{figure}[tbph]
\centering\includegraphics[width=\linewidth]{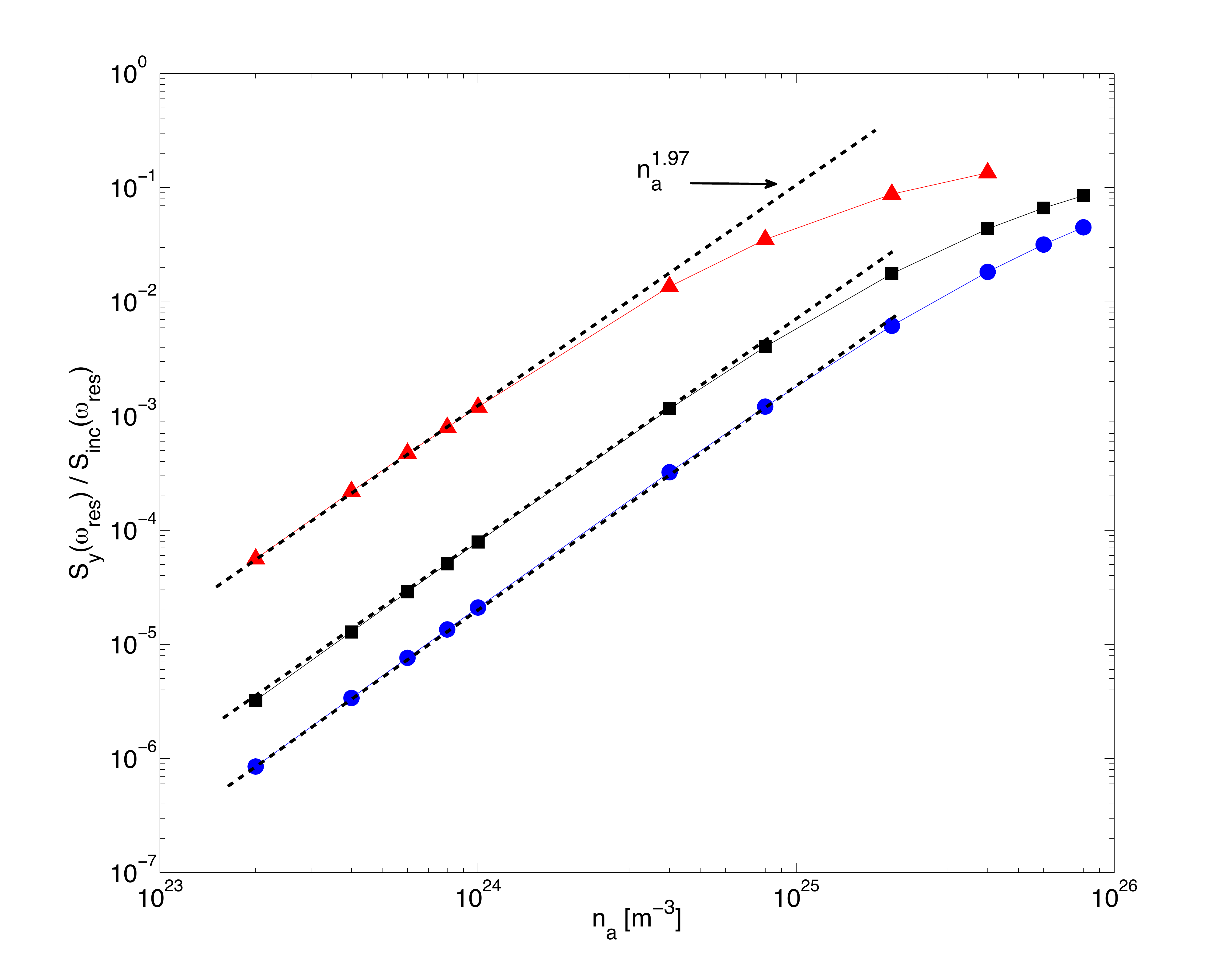}
\caption{(Color online). Scattering intensity (shown in double logarithmic scale) at the high frequency resonance, $\omega_{res}$, as a function of the atomic density, $n_a$, for three pure dephasing rates, $\gamma_p$ - blue circles: $\gamma_p=2\times10^{13}$ s$^{-1}$, black rectangles: $\gamma_p=10^{13}$ s$^{-1}$, and red triangles: $\gamma_p=2\times10^{12}$ s$^{-1}$. Other parameters are: $\omega_a=3.1$ eV, $\gamma_1=10^{12}$ s$^{-1}$, $R=25$ nm, $n_a=10^{26}$ m$^{-3}$, $\mu_{sp}=25$ Debye. The dashed straight lines represent fitting for each set of data demonstrating nearly ideal quadratic dependence on $n_a$ at low densities.}
\label{scaling}
\end{figure}

Fig. \ref{atoms} summarizes main results of our SPM calculations (note that we performed direct comparison of SPM data with that obtained via CW scheme for every set of parameters discussed below). First the dependence of the scattering intensity on the density of atoms in the cluster is depicted in Fig. \ref{atoms}A. The atomic transition frequency is fixed at $\omega_a=3.1$ eV (see figure caption for the rest of simulation parameters). The scattering radiation clearly exhibits two resonances; one, a relatively weak response, close to (slightly below) the atomic transition frequency. The other is a strong and broad peak at a higher frequency that moves to the blue at larger atomic densities. Additional simulations presented in Fig. \ref{scaling} show that the intensity of the high frequency mode (unlike the low frequency one) scales as $n_a^2$ for the low density of atoms suggesting a possible collective nature of the peak. It should be noted that this collective mode is noticeably wider than the atomic transition resonance.

The dependence of the scattering intensity on cluster's size is shown in Fig. \ref{atoms}B. The low energy resonance exhibits a red shift, when a radius of the cluster increases, while the resonant frequency of the high energy mode does not change with cluster's size. It, however, becomes significantly wider for larger clusters, which has been observed experimentally \cite{KobayashiCPL91}.

One of the advantages of the present calculations over the standard approach based on a dielectric model is the ability to examine the influence of the dephasing rate on optical properties. Fig. \ref{atoms}C shows simulation results obtained at three pure dephasing rates, $\gamma_{p}$, including the case without pure dephasing. We should note that at small $\gamma_{p}$ numerical simulations tend to become hard to converge at the frequencies near the collective resonance (in our case this occurs for $\gamma_{p}<4\times10^{12}$ s$^{-1}$). While for relatively high dephasing rates regular simulations require spatial steps, $\delta x$, on the order of 1 nm, the case without pure dephasing should be explored at $\delta x<0.2$ nm. It is interesting to note that the collective mode, while decreasing its width with the decrease of the dephasing rate, is still significantly wider than the atomic transition peak. In contrast, the low frequency peak becomes narrower with decreasing dephasing rate, with its width approaching $\gamma_1$. Note that the scattering actually shows a dip at the atomic frequency, with the low frequency peak slightly below it.

Fig. \ref{atoms}D explores how the scattering is affected by the matrix element of the atomic dipole moment, $\mu_{sp}$. Not surprisingly, the result for increasing $\mu_{sp}$ is qualitatively similar to that obtained with increasing atomic density. It is seen that larger $\mu_{sp}$ results in blue shift of the higher frequency collective mode, and in a red shift of the lower frequency mode. The former shows a quadratic dependence of the resonant frequency on the dipole moment, which has been theoretically discussed in the case of a sphere with uniformly distributed linear quantum dipoles \cite{PrasadGlauberPRA10}.

It is useful, for the sake of comparison, to consider the simplest theoretical description for the optical response of our system, by modeling it as a dielectric particle with a dielectric response function taken from the Clausius-Mossotti expression
\begin{equation}
\label{CM}
\varepsilon=\frac{1+2x}{1-x},
\end{equation}
where $x\equiv\frac{4\pi}{3}n\alpha$, $n$ is the number density of atoms, and $\alpha$ is the atomic polarizability. The absorption lineshape can be calculated \cite{Nitzan81} as the ratio between the dissipated power, $P_{diss}=\left( 1/2\right)\int d^3r\sigma \left| E\right|^2$ and the incident flux, $J_{\mbox{in}}=c\left| E_{\mbox{in}}\right|^2/\left(8\pi\right)$, where $\sigma=\left( \omega_{\mbox{in}} / \left( 4 \pi\right) \right)\Im\left(\varepsilon\right)$ is the conductivity, $\varepsilon$ is the dielectric response function, $E$ is the electric field in the particle, $E_{\mbox{in}}$ is the incident electric field and $c$ is the speed of light. For a small spherical particle \cite{Jackson}
\begin{equation}
\label{E-field}
E=\frac{3}{\varepsilon+2}E_{\mbox{in}}.
\end{equation}
Using these expressions we obtain the absorption cross-section of a small spherical particle of volume $\Omega$ in the form
\begin{equation}
\label{sigma}
\sigma_a=\frac{P_{diss}}{J_{\mbox{in}}}=\Omega\frac{\omega_{\mbox{in}}}{c}\left| \frac{3}{\varepsilon+2}\right|^2\Im\left(\varepsilon\right)=\frac{16\pi}{3}n\Omega\Im\left(\alpha\right).
\end{equation}
The imaginary part of the molecular optical polarizability is essentially a Lorentzian resonance peaked at the atomic transition frequency $\omega_a$. We see that the absorption cross-section in the Clausius-Mossotti approximation is proportional to the number of particles and to the absorption of a single particle, as would be predicted for a system of non-interacting particles. On the same level of theory, the dipole induced on the particle is \cite{Jackson}
\begin{equation}
\label{dipole}
\vec{\mu}=\frac{\varepsilon-1}{\varepsilon+2}\frac{3\Omega}{4\pi}\vec{E}_{\mbox{in}}=n\Omega\alpha\vec{E}_{\mbox{in}}.
\end{equation}
And, since the scattered light is proportional to $\left|\mu\right|^2$, it is predicted to go like the square of the number of particles, and to have a similar resonant behavior as a function of the incident frequency. Eqs. (\ref{sigma}) and (\ref{dipole}) describe essentially a system of non-interacting particles, occupying a volume with linear dimensions much smaller than the radiation wavelength, that respond coherently to the incident radiation. This approximation is valid at low atomic density. We will see below how dephasing and through-field interatomic interactions at higher densities affect this behavior.

While the calculation procedure applied here provides a route to explore the effect of dephasing on the optical response of atomic and molecular clusters, further studies will be needed in order to determine how much of this effect was indeed captured in our simulations. The destruction of phase is affected within our calculations on the mean field evolution of a single atom, feeling the effects of others through their mutual interaction with the radiation field. It is not obvious that this mean field implementation can capture the full physics of atoms going out of phase from each other. We leave this important technical question to a future study.

\begin{figure}[tbph]
\centering\includegraphics[width=\linewidth]{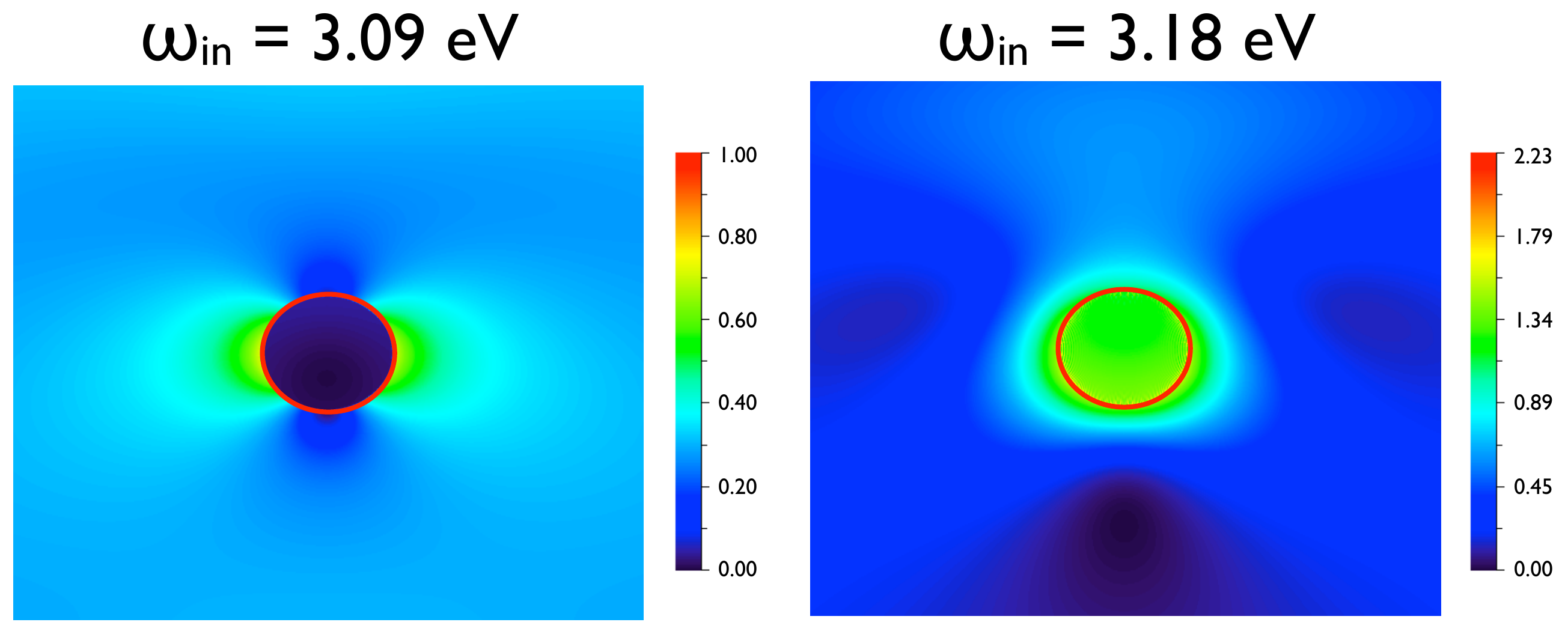}
\caption{(Color online). Spatial distributions of EM intensity (normalized to the incident intensity) in logarithmic scale at the low frequency resonance (left panel) and the high frequency resonance (right panel). Cluster's parameters are: $\omega_a=3.1$ eV, $\gamma_1=10^{12}$ s$^{-1}$, $\gamma_p=10^{13}$ s$^{-1}$, $R=25$ nm, $n_a=10^{26}$ m$^{-3}$, $\mu_{sp}=25$ Debye.}
\label{intensity}
\end{figure}

It is informative to explore the spatial dependence of EM intensity, $I\sim E_x^2+E_y^2$, at resonant conditions. Fig. \ref{intensity} shows intensity distributions calculated using steady-state solutions of Maxwell-Liouville equations for the two resonance modes. Clearly the EM intensity at the lower resonant frequency is mainly localized on the surface of the cluster exhibiting dipole radiation pattern similar to EM intensity distributions seen at the plasmon resonance for a single metal nanoparticle \cite{BarnesReview07}. The collective high frequency mode is distributed in the entire volume of the particle, where all atoms coherently participate in the radiation process. This suggests that the high frequency mode is more collective in nature than the low frequency one, consistent with its large density dependent shift from the atomic frequency and its $n_a^2$ scaling at low densities. The low frequency mode, involving fewer surface atoms, may be more atomic in nature. It is important to note however that this is not a single atom response since it clearly shifts with increasing atomic density.

\begin{figure}[tbph]
\centering\includegraphics[width=\linewidth]{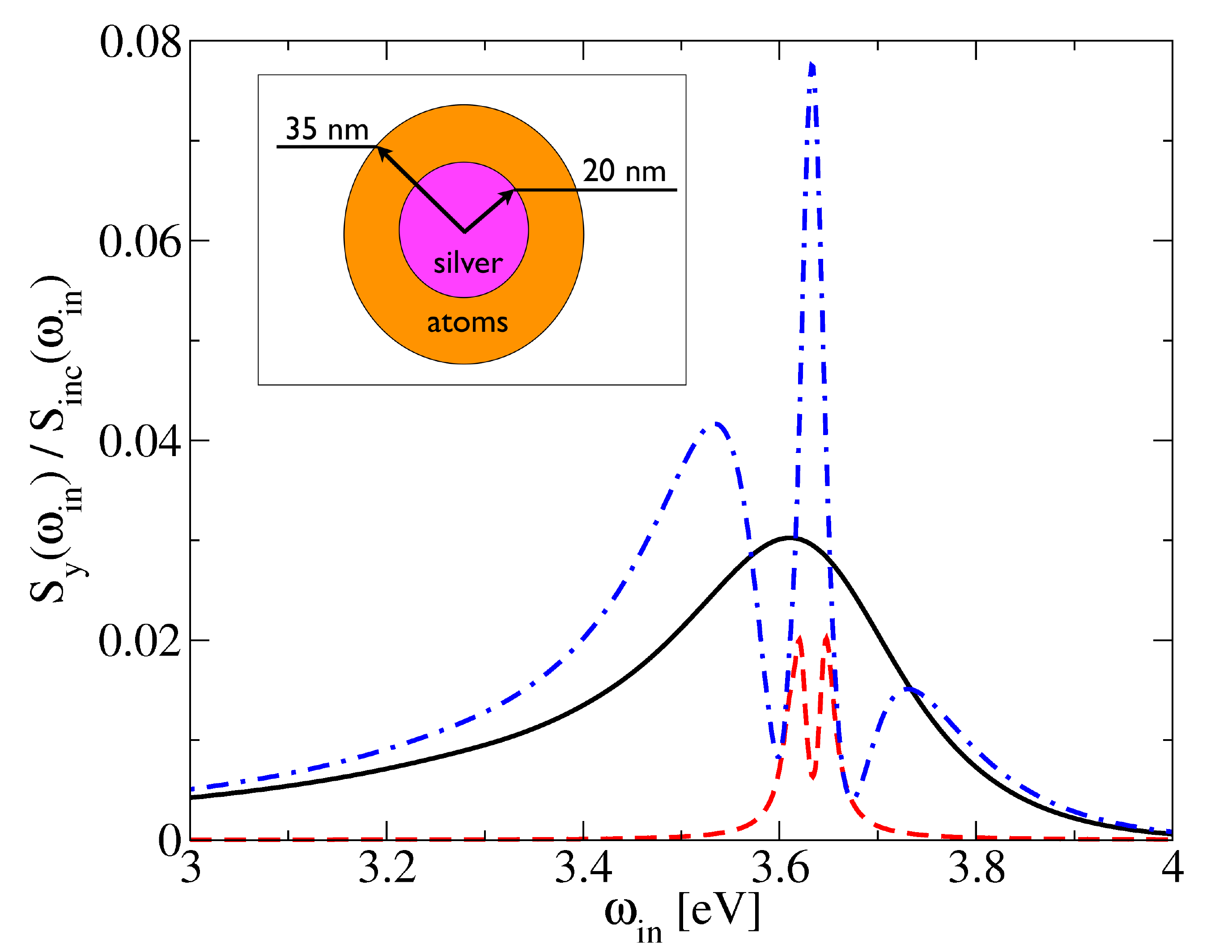}
\caption{(Color online). Scattering intensity as a function of the incident frequency for the core-shell particle shown in the inset. Black solid line shows the data for the silver nanoparticle without atomic shell, red dashed line presents simulations for the hollow atomic shell, and blue dash-dotted line demonstrates results obtained for atomic shell with a silver core. Parameters of the simulations are: $n_a=2.5\times10^{25}$ m$^{-3}$, $\omega_a=3.61$ eV, $\gamma_1=10^{12}$ s$^{-1}$, $\gamma_p=10^{13}$ s$^{-1}$, $\mu_{sp}=25$ Debye}
\label{core-shell}
\end{figure}

Optical properties of molecular aggregates resonantly coupled to plasmonic materials have been a subject of extensive research for the past several years. Sugawara et al. \cite{BartlettPRL06} demonstrated significant modification of transmission and reflection spectra of a gold film with deposited {\it J}-aggregates. It has been shown experimentally that SPP resonances notably affect molecular electronic structure leading to resonance splitting \cite{Ebbesen09}. The latter was proposed to be used for controlling optics of such hybrid material using femtosecond laser pulses \cite{Lienau10}. Moreover core-shell metal NPs with a shell comprised of optically active molecules have been recently studied experimentally \cite{Stadelmann10}. To demonstrate the generality of our approach we present simulations of the core-shell particle schematically depicted in the inset of Fig. \ref{core-shell}. Here a silver nanoparticle is shelled by a resonant atomic layer, with atomic transition frequency equaled to the SPP resonance of silver, $\omega_a=3.61$ eV. The optical properties of silver are described within the Drude model with parameters as in \cite{SukharevJPCA09}. The hollow atomic shell exhibits doubled collective mode (red dashed line in Fig. \ref{core-shell}), which corresponds to the symmetry of the problem and can be understood within the plasmon hybridization model proposed for noble metal core-shell particles \cite{NordlanderJCP04}. The important observation is a clear splitting of the SPP mode with additional strong peak centered near atomic transition frequency (blue dash-dotted line in Fig. \ref{core-shell}). The observed splitting as indicated in \cite{Ebbesen09} is due to the strong optical coupling of atoms with the SPP mode.

\section{Conclusion}
\label{conclusion}
We have presented a self-consistent electrodynamical model based on coupled Maxwell-Liouville equations that takes into account arbitrary polarization of the incident field. The proposed model is applied to investigate linear optical response of nanoscale atomic clusters in two dimensions merging classical electrodynamics with quantum mechanical description of atoms. The calculations can capture collective effects that play pivotal role in electrodynamics of nano-systems and, with limitations discussed below, includes the effect of dephasing on the optical response of these systems. 

We have found that spherical atomic clusters exhibit two well-distinguished resonances. The low energy resonance is close to the atomic transition frequency of individual atoms. The EM intensity distribution at this resonance is localized near the surface of a cluster. The high energy mode, where all atoms in the cluster coherently participate in the scattering, has clear collective nature. The dependence of the scattering intensity on various parameters was considered. It was demonstrated that the pure dephasing plays an important role in the scattering and absorption. Moreover we successfully applied our formalism to more complex systems, which comprise a resonant atomic shell and a silver core. 

Applications of the proposed scheme are many and vast. They range from complete three-dimensional description of nanoparticles resonantly coupled to ensembles of quantum particles, to nonlinear optical phenomena at the nanoscale. We note that our scheme can be extended to molecular systems, where one may investigate Raman processes. At the same time, we have indicated physically significant open technical issues. One is the limited ability of an approach based on classical electrodynamics to describe spontaneous emission and hence fluorescence. To account for such phenomena we need to modify the Maxwell-Liouville equations, Eqs.  (\ref{Maxwell}), (\ref{density matrix equations}), (\ref{polarization current}), so as to take into account the quantum nature of the radiation field, possibly using the quantization schemes described in Refs. \cite{Welsch1,Welsch2,Welsch3} or \cite{Tip1,Tip2} (which were shown to be equivalent \cite{Tip3}). Another is the need to examine the adequacy of mean field calculations of dephasing. All these will be subjects of our continuing studies.

\begin{acknowledgments}
The authors would like to acknowledge fruitful discussions with Dr. Maxim Efremov. M.S. is grateful to the financial and technical support from Arizona State University via startup funds. The research of A.N. is supported by the Israel Science Foundation, the Israel-US Binational Science Fundation, the European Science Council (FP7 /ERC grant no. 226628) and the Israel-Niedersachsen Research Fund.
\end{acknowledgments}

\newpage

\newpage

\newpage

\newpage

\newpage

\newpage


\begin{thebibliography}{99}
\bibitem{density note} Note that as in many studies of this type, a 2-dimensional solution does not correspond to a two dimensional system but to the 3-dimensional system, of infinite extent in the z-direction, obtained by displacing the xy cross-section along the z axis. The densities defined above remain per-unit-volume quantities.
\bibitem{CW volume}In the actual calculations $\rho_{22}$ and $\rho_{33}$ exhibit small low frequency oscillations that we eliminate by  time-averaging.
\bibitem{linear} We have numerically verified that all results of our simulations scale linearly with the incident intensity.
\end{thebibliography}
\end{document}